\documentclass[conference, final, letterpaper]{IEEEtran}
\usepackage[utf8]{inputenc}
\usepackage{graphicx}
\usepackage{stfloats}
\usepackage{amssymb}
\usepackage[cmex10]{amsmath}
\usepackage{amsthm} 
\usepackage{array}
\usepackage[usenames,dvipsnames]{xcolor}
\usepackage{hyperref}
\hypersetup{colorlinks=false, pdfborderstyle={/S/U/W 1},pdfborder=0 0 1, citebordercolor=Blue, filebordercolor=Red, linkbordercolor=Red, urlbordercolor=Blue}
\usepackage{url}
\usepackage{xparse}
\usepackage[english]{babel}
\usepackage[english = american]{csquotes}

\usepackage[backend=bibtex8, style=ieee-alex, url=false, isbn=false, doi=true, maxnames=5, firstinits=true]{biblatex}[2012/12/20 v2.4]
\AtEveryBibitem{
  \clearfield{month}
  \clearfield{edition}
  \clearname{editor}
}

\MakeOuterQuote{"}
\renewbibmacro{in:}{}
\DeclareListFormat{language}{}
\bibliography{qccarxiv}
\usepackage[ruled,vlined,titlenumbered]{algorithm2e}
\usepackage{tabularx}
\makeatletter
\newcommand{\removelatexerror}{\let\@latex@error\@gobble}
\makeatother
\newcommand{\printalgo}[1]
{
\begin{center}
\scalebox{0.97}{
\removelatexerror
\begin{algorithm}[H]
 #1
\end{algorithm}
}
\end{center}
}

\newtheorem{definition}{Definition}
\newtheorem{theorem}{Theorem}

\newtheorem{example}{Example}

\DeclareMathOperator{\diag}{diag}
\newcommand{\gcdab}[2]{\ensuremath{\gcd(#1,#2)}}

\DeclareMathOperator{\defi}{def}
\newcommand{\defeq}{\overset{\defi}{=}}
\newcommand{\defeqspace}{\overset{\hphantom{\defi}}{=}}
\newcommand{\defequiv}{\overset{\defi}{\equiv}}
\DeclareMathOperator{\rank}{rank}

\newcommand{\F}[1]{\mathbb F_{#1}}
\newcommand{\Fq}{\F{q}}
\newcommand{\Fxsub}[1]{\ensuremath{\mathbb{F}_{#1}[X]}}
\newcommand{\Fqx}{\Fxsub{q}}

\renewcommand{\vec}[1]{\mathbf #1}
\newcommand{\M}[2][\empty]{
  \ifthenelse{\equal{#1}{\empty}}
    {\ensuremath{\mathbf{#2}}}
    {\ensuremath{{\mathbf{#2}}_{#1}}}
}
\newcommand{\SET}[1]{\ensuremath{\mathsf{#1}}}
\newcommand{\LINq}[3]{\ensuremath{[#1,#2,#3]_{q}}}
\newcommand{\defset}[2][\empty]{
  \ifthenelse{\equal{#1}{\empty}}
    {\ensuremath{\SET{D}_{#2}}}
    {\ensuremath{\SET{D}^{[#1]}_{#2}}}
}
\newcommand{\CYC}{\ensuremath{\mathcal{C}}}
\newcommand{\HTconst}{\ensuremath{f}}

\newcommand{\lenseq}{\ensuremath{\delta}}
\newcommand{\noseq}{\ensuremath{\nu}}
\newcommand{\HTmult}{\ensuremath{z}}

\renewcommand{\tilde}{\widetilde}

\newcommand{\LINQCC}[5]{\ensuremath{[#1\cdot#2,#3,#4]_{#5}}}

\newcommand{\QCC}{\ensuremath{\mathcal{C}}}
\newcommand{\QCCk}{\ensuremath{k}} % dimension
\newcommand{\QCCd}{\ensuremath{d}} % minimum Hamming distance
\newcommand{\QCClen}{\ensuremath{m}}
\newcommand{\QCCcyc}{\ensuremath{\ell}}
\newcommand{\QCCext}{\ensuremath{r}}

\newcommand{\eigenvalue}[1]{\ensuremath{\lambda_{#1}}}
\DeclareDocumentCommand \eigenvector { ooo }
{
	\IfNoValueTF {#3}
	{  
		\IfNoValueTF {#2}
			{ % less then two values given
				\IfNoValueTF {#1}
				{ 	% NO parameter
					\ensuremath{\mathbf{v}}	
				}
				{ % one parameter given
					\ensuremath{v_{#1}}	
				}
			}
			{ % two values given
			\ensuremath{\mathbf{v}_{#1}^{\langle #2 \rangle}}			
			}
	}
	{% all three parameters are given
	\ensuremath{{v}_{#1,#3}^{\langle #2 \rangle}}
	}
}
\DeclareDocumentCommand \supp{o}
{
	\IfNoValueTF {#1}
	{  
		\ensuremath{\mathcal{Y}}
	}
	{
		\ensuremath{\mathcal{Y}_{#1}}
	}
}
\DeclareDocumentCommand \errorsup{o}
{
	\IfNoValueTF {#1}
	{  
		\ensuremath{\mathcal{E}}
	}
	{
		\ensuremath{\mathcal{E}_{#1}}
	}
}
\DeclareDocumentCommand \noerrors{o}
{
	\IfNoValueTF {#1}
	{  
		\ensuremath{\mathcal{\varepsilon}}
	}
	{
		\ensuremath{\mathcal{\varepsilon}_{#1}}
	}
}
\DeclareDocumentCommand \eigenspace{o}
{
	\IfNoValueTF {#1}
	{  
		\ensuremath{\mathcal{V}}
	}
	{
		\ensuremath{\mathcal{V}_{#1}}
	}
}
\newcommand{\eigencode}[1]{\ensuremath{\mathbb{C}(#1)}}
\newcommand{\mult}[1]{\ensuremath{u_{#1}}}
\newcommand{\ECn}{\ensuremath{n^{ec}}}
\newcommand{\ECk}{\ensuremath{k^{ec}}}
\newcommand{\ECd}{\ensuremath{d^{ec}}}
\newcommand{\HTbound}{\ensuremath{d^{\ast}}}
\newcommand{\interval}[1]{\ensuremath{[#1]}}
\newcommand\bigzero{\makebox(0,0){\text{\huge0}}}
\begin{document}
\title{Decoding of Quasi-Cyclic Codes up to A New Lower Bound on the Minimum Distance}

\IEEEoverridecommandlockouts

\author{\IEEEauthorblockN{Alexander Zeh}\thanks{A. Zeh has been supported by the German research council (Deutsche Forschungsgemeinschaft, DFG) under grant Ze1016/1-1. S. Ling has been supported by NTU Research Grant M4080456.}
\IEEEauthorblockA{Computer Science Department\\
Technion---Israel Institute of Technology\\
Haifa, Israel\\
\texttt{alex@codingtheory.eu}
}
\and
\IEEEauthorblockN{San Ling}
\IEEEauthorblockA{Division of Mathematical Sciences, School of Physical \& \\
Mathematical Sciences, Nanyang Technological University\\
Singapore, Republic of Singapore\\
\texttt{lingsan@ntu.edu.sg}}
}

\maketitle

\begin{abstract}
A new lower bound on the minimum Hamming distance of linear quasi-cyclic codes over finite fields is proposed. It is based on spectral analysis and generalizes the Semenov--Trifonov bound in a similar way as the Hartmann--Tzeng bound extends the BCH approach for cyclic codes. Furthermore, a syndrome-based algebraic decoding algorithm is given.
\end{abstract}

\begin{IEEEkeywords}
Bound on the minimum distance, efficient decoding, quasi-cyclic code, spectral analysis
\end{IEEEkeywords}

\section{Introduction}
The class of linear quasi-cyclic codes over finite fields is a generalization of cyclic codes and is known to be asymptotically good (see, e.g., Chen--Peterson--Weldon~\cite{chen_results_1969}).  Many of the best known linear codes belong to this class (see, e.g., Gulliver--Bhargava~\cite{gulliver_best_1991} and Chen's database~\cite{chen_database_2014}). Several good LDPC codes are quasi-cyclic and the connection to convolutional codes was investigated among others in~\cite{solomon_connection_1979,esmaeili_link_1998,lally_algebraic_2006}.

The algebraic structure of quasi-cyclic codes was exploited in various ways (see, e.g., Lally--Fitzpatrick~\cite{lally_algebraic_2001}, Ling--Solé~\cite{ling_algebraic_2001, ling_algebraic_2003, ling_algebraic_2005}, Barbier~\textit{et al.}~\cite{barbier_quasi-cyclic_2012,barbier_decoding_2013}), but the estimates on the minimum distance are far away from the real minimum distance and thus the guaranteed decoding radius.
Recently, Semenov and Trifonov~\cite{semenov_spectral_2012} developed a spectral analysis of quasi-cyclic codes based on the work of Lally and Fitzpatrick~\cite{lally_construction_1999, lally_algebraic_2001} and formulated a BCH-like lower bound on the minimum distance of quasi-cyclic codes.

We generalize the Semenov--Trifonov~\cite{semenov_spectral_2012} bound on the minimum distance of quasi-cyclic codes. Our new approach is similar to the Hartmann--Tzeng (HT,~\cite{hartmann_decoding_1972, hartmann_generalizations_1972}) bound, which generalizes the BCH~\cite{bose_class_1960,hocquenghem_codes_1959} bound for cyclic codes. Moreover, we prove a quadratic-time syndrome-based algebraic decoding algorithm up to the new bound and show that it is advantageous in the case of burst errors.

This paper is organized as follows. In Section~\ref{sec_Preliminaries}, we recall the Gröbner basis representation of quasi-cyclic codes of Lally--Fitzpatrick~\cite{lally_construction_1999,lally_algebraic_2001} and the definitions of the spectral method of Semenov--Trifonov~\cite{semenov_spectral_2012}.
The new HT-like bound on the minimum distance is formulated and proven in Section~\ref{sec_HartmannTzeng}. Section~\ref{sec_Decoding} describes a syndrome-based decoding algorithm up to our bound and shows that in the case of burst errors more symbol errors can be corrected. We draw some conclusions in Section~\ref{sec_conclusion}.

\section{Preliminaries} \label{sec_Preliminaries}
\subsection{Reduced Gröbner Basis}
Let $\Fq$ denote the finite field of order $q$ and $\Fqx$ the polynomial ring over $\Fq$ with indeterminate $X$. Let $z$ be a positive integer and denote by $\interval{z}$ the set of integers $\{0,1,\dots,z-1\}$. A vector of length $n$ is denoted by a lowercase bold letter as $\vec{v} = (v_0 \, v_1 \, \dots \, v_{n-1})$ and $\vec{v} \circ \vec{w}$ denotes the scalar product $\sum_{i=0}^{n-1} v_i w_i$ of two vectors $\vec{v}, \vec{w}$ of length $n$.
An $m \times n$ matrix is denoted by a capital bold letter as $\M{M}=(m_{i,j})_{i \in \interval{m}}^{j \in \interval{n}}$. 

A linear \LINQCC{\QCClen}{\QCCcyc}{\QCCk}{\QCCd}{q} code $\QCC$ of length $\QCClen \QCCcyc$, dimension $\QCCk$ and minimum Hamming distance $\QCCd$ over $\Fq$ is $\QCCcyc$-quasi-cyclic if every cyclic shift by $\QCCcyc$ of a codeword is again a codeword of $\QCC$, more explicitly if:
\begin{align*}
&(c_{0,0} \dots c_{\QCCcyc-1,0} \ c_{0,1} \dots c_{\QCCcyc-1,1} \ \dots \ c_{\QCCcyc-1,\QCClen-1}) \in \QCC \Rightarrow \\
& (c_{0,\QCClen-1} \dots c_{\QCCcyc-1,\QCClen-1} \ c_{0,0} \dots c_{\QCCcyc-1,0} \ \dots c_{\QCCcyc-1,\QCClen-2}) \in \QCC.
\end{align*}
We can represent a codeword of an \LINQCC{\QCClen}{\QCCcyc}{\QCCk}{\QCCd}{q} $\QCCcyc$-quasi-cyclic code as $\mathbf{c}(X) = (c_0(X) \ c_1(X) \ \dots \ c_{\QCCcyc-1}(X)) \in \Fqx^{\ell} $, where
\begin{equation*}
c_i(X) \defeq \sum_{j=0}^{\QCClen-1} c_{i,j} X^{j}, \quad \forall i \in \interval{\QCCcyc}.
\end{equation*}
Then, the defining property of $\QCC$ is that each component $c_i(X)$ of $\mathbf{c}(X)$ is closed under multiplication by $X$ and reduction modulo $X^{\QCClen}-1$.
Lally and Fitzpatrick~\cite{lally_construction_1999, lally_algebraic_2001} showed that this enables us to see a quasi-cyclic code as an $\Fqx/\langle X^{\QCClen}-1 \rangle$-submodule of the algebra $(\Fqx/\langle X^{\QCClen}-1 \rangle)^{\QCCcyc}$ and they proved that every quasi-cyclic code has a generating set in the form of a reduced Gröbner basis with respect to the position-over-term order in $\Fqx^{\QCCcyc}$.
This basis can be represented in the form of an upper-triangular $\ell \times \ell$ matrix with entries in $\Fqx$ as follows: 
\begin{equation} \label{def_GroebBasisMatrix}
\mathbf{\tilde{G}}(X) =
\begin{pmatrix}
g_{0,0}(X) & g_{0,1}(X) & \cdots  & g_{0,\QCCcyc-1}(X) \\
 & g_{1,1}(X) & \cdots & g_{1,\QCCcyc-1}(X) \\
\multicolumn{2}{c}{\bigzero}& \ddots & \vdots \\
 &  & & g_{\QCCcyc-1,\QCCcyc-1}(X)
\end{pmatrix},
\end{equation}
where the following conditions must be fulfilled:
\begin{tabular}[htb]{lll}
1) & $g_{i,j}(X) = 0,$ & $\forall 0 \leq j < i < \QCCcyc$, \\
2) & $\deg g_{j,i}(X) < \deg g_{i,i}(X),$ & $ \forall j < i, i \in \interval{\QCCcyc}$,\\
3) & $g_{i,i}(X) | (X^{\QCClen}-1),$ & $\forall i \in \interval{\QCCcyc}$,\\
4) & if $g_{i,i}(X)=X^{\QCClen}-1$ then \\ 
& $g_{i,j}(X)=0,$ & $ \forall j=i+1,\dots,\QCCcyc-1$.
\end{tabular}
A codeword of $\QCC$ can be represented as $\mathbf{c}(X) = \mathbf{a}(X) \mathbf{\tilde{G}}(X)$ and it follows that $\QCCk = \QCClen \QCCcyc - \sum_{i=0}^{\QCCcyc-1} \deg g_{i,i}(X)$.

For $\QCCcyc=1$, the generator matrix $\mathbf{\tilde{G}}(X)$ becomes the well-known generator polynomial of a cyclic code of degree $\QCClen-\QCCk$. We restrict ourselves throughout this paper to the single-root case, i.e., $\gcd(\QCClen, \text{char}(\Fq))=1$.

\subsection{Spectral Analysis of Quasi-Cyclic Codes}
Let $\mathbf{\tilde{G}}(X)$ be the upper-triangular generator matrix of a given \LINQCC{\QCClen}{\QCCcyc}{\QCCk}{\QCCd}{q} $\QCCcyc$-quasi-cyclic code $\QCC$ in reduced Gröbner basis form as in~\eqref{def_GroebBasisMatrix}. Let $\alpha \in \F{q^{\QCCext}}$ be an $\QCClen$-th root of unity.
An eigenvalue $\eigenvalue{i} = \alpha^{j_i}$ of $\QCC$ is defined to be a root of $\det(\mathbf{\tilde{G}}(X))$, i.e., a root of $\prod_{i=0}^{\QCCcyc-1} g_{i,i}(X)$.
The \textit{algebraic} multiplicity of $\eigenvalue{i}$ is the largest integer $ \mult{i} $ such that 
$(X-\eigenvalue{i})^{\mult{i}} \mid \det(\mathbf{\tilde{G}}(X)).$
Semenov and Trifonov~\cite{semenov_spectral_2012} defined the \textit{geometric} multiplicity of an eigenvalue $\eigenvalue{i}$ as the dimension of the right kernel of the matrix $\mathbf{\tilde{G}}(\eigenvalue{i})$, i.e., the dimension of the solution space of the homogeneous linear system of equations:
\begin{equation} \label{eq_eigenvectors}
\mathbf{\tilde{G}}(\eigenvalue{i}) \eigenvector =  \textbf{0}.
\end{equation}
The solution space of~\eqref{eq_eigenvectors} is called the right kernel eigenspace and it is denoted by $\eigenspace[i]$. Furthermore, it was shown that, for a matrix $\mathbf{\tilde{G}}(X) \in \Fqx^{\QCCcyc \times \QCCcyc}$ in the reduced Gröbner basis representation, the algebraic multiplicity $\mult{i}$ of an eigenvalue $\eigenvalue{i}$ equals the geometric multiplicity (see~\cite[Lemma 1]{semenov_spectral_2012}).
Moreover, they gave in~\cite{semenov_spectral_2012} an explicit construction of the parity-check matrix of an \LINQCC{\QCClen}{\QCCcyc}{\QCCk}{\QCCd}{q} $\QCCcyc$-quasi-cyclic code $\QCC$ and proved a BCH-like~\cite{bose_class_1960, hocquenghem_codes_1959} lower bound on $\QCCd$ using the parity-check matrix and the so-called eigencode. We generalize their approach, but do not explicitly need the parity-check matrix for the proof though the eigencode is still needed.
\begin{definition}[Eigencode] \label{def_eigencode}
Let $\eigenspace{} \subseteq \F{q^{\QCCext}}^{\QCCcyc}$ be an eigenspace. Define the $\LINq{\ECn=\QCCcyc}{\ECk}{\ECd}$ eigencode corresponding to $\eigenspace{}$ by
\begin{equation} \label{def_eq_eigencode}
\eigencode{\eigenspace{}} \defeq \left \lbrace (c_0 \dots c_{\QCCcyc-1}) \in \Fq^{\QCCcyc} \mid \forall \eigenvector \in \eigenspace : \sum_{i=0}^{\QCCcyc-1} \eigenvector[i] c_i = 0  \right\rbrace.
\end{equation}
\end{definition}
If there exists $\eigenvector = (\eigenvector[0] \ \eigenvector[1] \  \dots \ \eigenvector[\QCCcyc-1]) \in \eigenspace$ such that the elements $\eigenvector[0], \eigenvector[1], \dots, \eigenvector[\QCCcyc-1] $ are linearly independent over $\Fq$, then $\eigencode{\eigenspace{}} = \{ (0 \ 0 \ \dots \ 0) \} $ and $\ECd$ is infinity.
To  describe quasi-cyclic codes explicitly, we need to recall the following facts about cyclic codes. A $q$-cyclotomic coset $M_i$ is defined as:
\begin{equation} \label{eq_cyclotomiccoset}
 M_i \defeq \Big\{ iq^j \mod \QCClen \, \vert \, j \in \interval{a} \Big\},
\end{equation}
where $a$ is the smallest positive integer such that $iq^{a} \equiv i \bmod \QCClen$. 
The minimal polynomial in $\Fqx$ of the element $\alpha^i \in \F{q^{\QCCext}}$ is given by $m_i(X) = \prod_{j \in M_i} (X-\alpha^j)$.

\section{Improved Lower Bound} \label{sec_HartmannTzeng}
In this section, we generalize the lower bound on the minimum distance of quasi-cyclic codes given in~\cite[Thm.~2]{semenov_spectral_2012} in a similar way as the Hartmann--Tzeng bound~\cite{hartmann_decoding_1972, hartmann_generalizations_1972} generalizes the BCH bound~\cite{bose_class_1960, hocquenghem_codes_1959} for cyclic codes.
\begin{theorem}[New Lower Bound] \label{theo_HTBound}
Let $\QCC$ be an \LINQCC{\QCClen}{\QCCcyc}{\QCCk}{\QCCd}{q} $\QCCcyc$-quasi-cyclic code and let $\alpha \in \F{q^{\QCCext}}$ denote an element of order $\QCClen$.
Define the set 
\begin{align*}
D \defeq \big\{ \HTconst, \HTconst+\HTmult,\dots, & \HTconst+(\lenseq-2)\HTmult, \\ 
    \HTconst+1,\HTconst+&1+\HTmult,\dots,\HTconst+ 1+(\lenseq -2)\HTmult, \\
     \ddots & \qquad \ddots \qquad  \cdots \qquad \ddots \\
   \HTconst & + \noseq,\HTconst+\noseq+\HTmult ,\dots,\HTconst +\noseq+(\lenseq-2)\HTmult \big\},
\end{align*}
for some integers $\HTconst$, $\lenseq > 2 $ and $\HTmult > 0$ with $\gcdab{\QCClen}{\HTmult} = 1$. Let the eigenvalues $\eigenvalue{i} = \alpha^i, \forall i \in D$, their corresponding eigenspaces $\eigenspace[i], \forall i \in D$, be given, and let their intersection be $\eigenspace \defeq \bigcap_{i \in D} \eigenspace[i]$.

Let $\ECd$ denote the distance of the eigencode $\eigencode{\eigenspace{}}$ and let $\eigenvector= (\eigenvector[0] \ \eigenvector[1] \  \dots \ \eigenvector[\QCCcyc-1]) \in \eigenspace$ be an eigenvector where $\eigenvector[0], \eigenvector[1], \dots, \eigenvector[\QCCcyc-1]$ are linearly independent over $\Fq$. If  
\begin{align} \label{eq_HTBound}
\sum_{i=0}^{\infty} \mathbf{c}(\alpha^{\HTconst+\HTmult i + j }) \circ \eigenvector X^i \equiv 0 \mod X^{\lenseq-1}, \; \forall  j \in \interval{\noseq+1},
\end{align}
holds for all $\mathbf{c}(X) = \big( c_0(X) \ c_1(X) \ \dots \ c_{\QCCcyc-1}(X) \big) \in \CYC$, then, $d \geq \HTbound \defeq \min(\lenseq+\noseq, \ECd)$.
\end{theorem}
\begin{IEEEproof}
Let $c_i(X) = \sum_{j \in \supp[i]}c_{i,j}X^j, \forall i \in \interval{\QCCcyc}$, where $c_{i,j} \in \Fq$.
We can write the LHS of~\eqref{eq_HTBound} more explicitly:
\begin{align} \label{eq_LHSb}
& \sum_{i=0}^{\infty} \Bigg(\sum_{t=0}^{\QCCcyc-1} c_{t}(\alpha^{\HTconst+\HTmult i + j })  \eigenvector[t] \Bigg) X^i \equiv 0 \bmod X^{\lenseq-1}, \forall j \in \interval{\noseq+1}.
\end{align}
Now, define:
\begin{align} \label{eq_UnionOfWeights}
\supp & = \{i_0,i_1,\dots,i_{y-1} \} \defeq  \bigcup_{i=0}^{\QCCcyc-1} \supp[i] \subseteq \interval{\QCClen}. 
\end{align}
We obtain from~\eqref{eq_LHSb} with~\eqref{eq_UnionOfWeights} :
\begin{align} \label{eq_ExplCodewords}
\sum_{i=0}^{\infty} \Bigg( \sum_{s \in \supp} \Bigg( \sum_{t=0}^{\QCCcyc-1} c_{t,s} \eigenvector[t] \Bigg) & \alpha^{(\HTconst+\HTmult i + j)s } \Bigg)  X^i \nonumber \\
& \equiv 0 \bmod X^{\lenseq-1}, \; \forall j \in \interval{\noseq+1}.
\end{align}
We define $\QCClen$ elements in $\F{q^{\QCCext}}$ as follows:
\begin{equation} \label{eq_BigElement}
C_s \defeq \sum_{t=0}^{\QCCcyc-1} c_{t,s} \eigenvector[t], \quad \forall s \in \interval{\QCClen}.
\end{equation}
With~\eqref{eq_BigElement}, we can simplify~\eqref{eq_ExplCodewords} to
\begin{align} \label{eq_ExplCodewordsLarge}
\sum_{i=0}^{\infty} \Bigg( \sum_{s \in \supp}  C_s & \alpha^{(\HTconst+\HTmult i + j)s }  \Bigg)  X^i \equiv 0 \bmod X^{\lenseq-1}, \forall j \in \interval{\noseq+1}.
\end{align}
We linearly combine the $\noseq+1$ sequences of~\eqref{eq_ExplCodewordsLarge}, multiply each of them by an element $\omega_j \in \F{q^{\QCCext}} \backslash \{0\}$ and obtain:
\begin{equation} \label{eq_LinerCombineda}
\sum_{j = 0}^{\noseq} \omega_j \sum_{i=0}^{\infty} \Bigg( \sum_{s \in \supp} C_{s}\alpha^{(\HTconst+\HTmult i+j)s} \Bigg) X^i \equiv 0 \bmod X^{\lenseq-1}.
\end{equation}
Interchanging the sums in~\eqref{eq_LinerCombineda} leads to:
\begin{equation} \label{eq_ExpressionBeforeVandermonde}
\sum_{i=0}^{\infty} \sum_{s \in \supp} \Big( C_{s}\alpha^{(\HTconst+\HTmult i)s} \sum_{j = 0}^{\noseq}  \omega_j \alpha^{j s} \Big) X^i \equiv 0 \bmod X^{\lenseq-1}.
\end{equation}
We choose $\omega_0,\omega_1, \dots, \omega_{\noseq}$ such that the first $\noseq$ terms with coefficients $C_{i_0}, C_{i_1}, \dots, C_{i_{\noseq-1}}$ are annihilated. We obtain the following linear $(\noseq+1) \times (\noseq+1)$ system of equations:
\begin{align} \label{eq_SystemForCoefficients}
\begin{pmatrix}
1 & \alpha^{i_0} & \alpha^{i_0 2} & \cdots & \alpha^{i_0 \noseq}  \\ 
1 & \alpha^{i_1} & \alpha^{i_1 2} & \cdots & \alpha^{i_1 \noseq }  \\ 
\vdots &  \vdots & \vdots & \ddots  &  \vdots\\ 
1 & \alpha^{i_{\noseq}} & \alpha^{i_{\noseq} 2} & \cdots & \alpha^{i_{\noseq} \noseq}  \\ 
\end{pmatrix}
\begin{pmatrix}
\omega_0 \\ 
\omega_1 \\ 
\vdots \\
\omega_{\noseq}
\end{pmatrix}
=
\begin{pmatrix}
0 \\ 
\vdots \\
0 \\ 
1
\end{pmatrix},
\end{align}
with Vandermonde structure and therefore the non-zero solution is unique.
Let $\tilde{\mathcal{Y}} \defeq \mathcal{Y} \setminus \{ i_0,i_1,\dots, i_{\noseq-1}\}$. Then we can rewrite~\eqref{eq_ExpressionBeforeVandermonde}:
\begin{equation} \label{eq_BeforeGeometric}
\sum_{i=0}^{\infty}  \sum_{s \in \tilde{\mathcal{Y}}} \Big( C_{s} \alpha^{(\HTconst+\HTmult i)s} \sum_{j = 0}^{\noseq}  \omega_j \alpha^{j s} \Big) X^{i} \equiv 0 \bmod X^{\lenseq-1}.
\end{equation}
With the geometric series we get from~\eqref{eq_BeforeGeometric}:
\begin{align*}
\sum_{s \in \tilde{\mathcal{Y}}} \frac{C_{s} \alpha^{s \HTconst}
\Big(\sum\limits_{j = 0}^{\noseq}  \omega_j \alpha^{j s} \Big) }{1-\alpha^{\HTmult s} X} & \equiv 0 \mod X^{\lenseq-1},
\end{align*}
and writing each fraction as an equivalent fraction with the least common denominator leads to:
\begin{equation} \label{eq_FinalExpressionHT}
\frac{\sum\limits_{s \in \tilde{\mathcal{Y}}}  \Big( C_{s} \alpha^{s \HTconst} \big( \sum\limits_{j = 0}^{\noseq}  \omega_j \alpha^{j s} \big)  \prod\limits_{\substack{h \in \tilde{\mathcal{Y}}\\ h \neq s}}  (1-\alpha^{\HTmult h} X) \Big) }{\prod\limits_{s \in \tilde{\mathcal{Y}}} (1-\alpha^{\HTmult s} X)} \equiv 0  \bmod X^{\lenseq-1}, 
\end{equation}
where the degree of the numerator is at most $|\tilde{\mathcal{Y}}|-1 = y-\noseq-1$ and has to be at least $\lenseq-1$.

To bound the distance $\QCCd$ we distinguish two cases. For the first case where $\ECd > \lenseq + \noseq$, at least $y-\noseq$ elements $C_i \in \F{q^{\QCCext}}$ have to be non-zero such that \eqref{eq_FinalExpressionHT} holds, i.e., at least $y-\noseq$ elements $c_{t_0,i_0},c_{t_1,i_1}, \dots, c_{t_{y-1-\noseq},i_{y-1-\noseq}} \in \Fq$ for $t_0, \dots, t_{y-1-\noseq} $ distinct, have to be non-zero and therefore $\QCCd-\noseq-1 \geq \lenseq -1  \Longleftrightarrow \QCCd \geq \lenseq+\noseq$.
For the second case where $\ECd < \lenseq + \noseq$, at least $\ECd$ elements $c_{j,i_0}, c_{j,i_1}, \dots, c_{j,i_{\ECd-1}}$ have to be non-zero (see~\eqref{eq_BigElement}) such that $C_j = 0$ and if all the other $C_s, s \in \tilde{\mathcal{Y}} \backslash \{j\}$, are zero, then the LHS of \eqref{eq_FinalExpressionHT} becomes zero. In this case $\QCCd \geq \ECd$.
\end{IEEEproof}
For $\noseq = 0$, the bound of Theorem~\ref{theo_HTBound} becomes the bound of Semenov--Trifonov (see~\cite[Thm. 2]{semenov_spectral_2012}). 
We chose to state Thm.~\ref{theo_HTBound} in terms of all $\mathbf{c}(X) \in \QCC$ (see~\eqref{eq_HTBound}) to easily obtain a syndrome expression (see Section~\ref{sec_Decoding}). 
In practice, from the spectral analysis of $\tilde{{\mathbf G}}(X)$, one can search for eigenvalues of the form $\alpha^i$, for $i$ in some $D$ of the form in Thm.~\ref{theo_HTBound}, and determine the corresponding eigencode with its minimum distance. The condition~\eqref{eq_HTBound} is then automatically satisfied for all codewords $\mathbf{c}(X) \in \QCC$, with the corresponding $\HTconst$, $\HTmult$ and $\lenseq$.
\begin{example}[HT-like Bound for Quasi-Cyclic Code] \label{ex_HTbound}
Let $\QCC$ be the binary $\LINQCC{63}{2}{100}{6}{2}$ $2$-quasi-cyclic code with $2 \times 2$ generator matrix in reduced Gröbner form as defined in~\eqref{def_GroebBasisMatrix}:
\begin{equation*}
\mathbf{\tilde{G}}(X) = 
\begin{pmatrix}
g_{0,0}(X) & g_{0,1}(X) \\
0 & g_{1,1}(X)
\end{pmatrix},
\end{equation*}
where:
\begin{align*}
g_{0,0}(X) & = m_0(X)m_1(X)m_9(X),\\
g_{0,1}(X) & = g_{0,0}(X) a_{0,1}(X), \quad g_{1,1}(X)  = g_{0,0}(X) m_5(X),
\end{align*}
and $a_{0,1}(X)= X^4+X^3+X^2+X+1$ with $\deg a_{0,1}(X) < \deg m_5(X)$ and $ a_{0,1}(X) \nmid (X^{63}-1)$. 

Let $\alpha \in \F{2^6} \cong \Fxsub{2}/(X^6 + X^4 + X^3 + X + 1)$ be an element of order $63$. The eigenvalues $\eigenvalue{i} = \alpha^i, i \in \{0, 1, 2, 4, 8, 9, 16, 18, 32, 36 \} = M_0 \cup M_1 \cup M_9$ are the roots of $g_{0,0}(X)$, $g_{0,1}(X)$, $g_{1,1}(X)$ and have (algebraic and geometric) multiplicity two. Therefore, the corresponding eigenvectors span the full space $\F{2^{6}}^{2}$. The distinct eigenvectors $\eigenvector^{(i)}, \forall i \in M_5$, are in $\F{2^{6}}^{2}$ and $\eigenvector[0]^{(i)}, \eigenvector[1]^{(i)} \in \F{2^{6}}$, are linearly independent over $\F{2}$ for each $i \in M_5$.

With $\HTconst = 0$, $\HTmult = 4$, $\lenseq = 4$, $\noseq = 1 $, we obtain two consecutive sequences of eigenvalues $\alpha^0,\alpha^4,\alpha^8$ and $\alpha^1,\alpha^5,\alpha^9$ of length three, where $\eigenvector[0]^{(5)}=1, \eigenvector[1]^{(5)}=\alpha^4+1$, are linearly independent over $\F{2}$ and $\eigenvector^{(5)}$ is contained in the intersection of the eigenspaces $\eigenspace[i], i \in D \defeq \{0,4,8,1,5,9\}$, and therefore $\ECd = \infty$ of $\eigencode{\cap_{i \in D} \eigenspace[i]}$. With Theorem~\ref{theo_HTBound}, we can bound $\QCCd$ to be at least $\lenseq+\noseq = 5$, which is one less than the actual minimum distance for the $\LINQCC{63}{2}{100}{6}{2}$ $2$-quasi-cyclic code. The bound of Semenov--Trifonov gives $\QCCd \geq 4$.
\end{example}

\section{Syndrome-Based Decoding of Quasi-Cyclic Codes} \label{sec_Decoding}
In this section, we develop a syndrome-based decoding algorithm, which guarantees to correct up to $\lfloor (\HTbound-1)/2 \rfloor$ symbol errors in $\Fq$. Let the received word of a given $\LINQCC{\QCClen}{\QCCcyc}{\QCCk}{\QCCd}{q}$ $\QCCcyc$-quasi-cyclic code be:
\begin{align*}
\mathbf{r}(X) & = \big( r_0(X) & \ \dots \ r_{\QCCcyc-1}(X) & \big) \\
 &  = \big( c_0(X) + e_0(X) &  \ \dots \ c_{\QCCcyc-1}(X) & + e_{\QCCcyc-1}(X) \big),
\end{align*}
where
\begin{equation} \label{eq_errorword}
e_i(X) = \sum_{j \in \errorsup[i]} e_{i,j} X^j, \quad i \in \interval{\QCCcyc},
\end{equation}
are $\QCCcyc$ error polynomials in $\Fqx$ with $\noerrors[i] \defeq |\errorsup[i]|$ and degree less than $\QCClen$. The number of errors in $\Fq$ is $\tilde{\noerrors} \defeq \sum_{i=0}^{\QCCcyc-1} \noerrors[i]$. Define the following set of burst errors: 
\begin{equation} \label{eq_ModError}
\errorsup \defeq \bigcup_{i=0}^{\QCCcyc-1} \errorsup[i] \subseteq \interval{\QCClen}. 
\end{equation} 
with cardinality $\noerrors \defeq  |\errorsup| \leq \tilde{\noerrors}$.

In the following, we describe a decoding procedure that is able to decode up to $\noerrors \leq \tau$ errors, where:
\begin{equation} \label{eq_DecRadius}
\tau \leq \frac{\HTbound-1}{2}.
\end{equation}

Let $\alpha \in \F{q^{\QCCext}}$ denote an $\QCClen$-th root of unity and let the $(\noseq+1)(\lenseq-1)$ eigenvalues $\eigenvalue{i} = \alpha^{\HTconst+i \HTmult+j}, \forall i \in \interval{\lenseq-1}, j \in \interval{\noseq+1}$, the integer $\HTconst$ and the integer $\HTmult > 0$ with $\gcd(\HTmult,\QCClen)=1$ be given as stated in Thm.~\ref{theo_HTBound}. Furthermore, let $\eigenspace = \bigcap_{i \in \interval{\lenseq-1}, j \in \interval{\noseq+1}} \eigenspace[\HTconst + i \HTmult + j]$ and let one eigenvector $\eigenvector = (\eigenvector[0] \ \eigenvector[1] \ \dots \ \eigenvector[\QCCcyc-1]) \in \eigenspace$, where $\eigenvector[0], \eigenvector[1], \dots, \eigenvector[\QCCcyc-1] $ are linearly independent over $\Fq$, be given. We assume that the minimum distance of the corresponding eigencode $\eigencode{\eigenspace}$ is greater than $\lenseq+\noseq$. Then, we define the following $\noseq+1$ syndrome polynomials in $\Fxsub{q^{\QCCext}}$: 
\begin{align} \label{eq_DefSyndromes}
S_{t}(X) & \defequiv  \sum_{i=0}^{\infty} \Bigg( \sum_{j=0}^{\QCCcyc-1} r_j(\alpha^{\HTconst +i \HTmult+t}) \eigenvector[j] \Bigg) X^i \mod X^{\lenseq-1} \nonumber \\
& \defeqspace  \sum_{i=0}^{\lenseq-2} \Bigg( \sum_{j=0}^{\QCCcyc-1} r_j(\alpha^{\HTconst +i \HTmult+t}) \eigenvector[j] \Bigg) X^i, \: \forall t \in \interval{\noseq+1}.
\end{align}
From Thm.~\ref{theo_HTBound} it follows that the syndrome polynomials as defined in~\eqref{eq_DefSyndromes} depend only on the error and therefore:
\begin{align*}
S_{t}(X) & =  \sum_{i=0}^{\lenseq-2} \Bigg( \sum_{j=0}^{\QCCcyc-1} e_j(\alpha^{\HTconst +i \HTmult+t}) \eigenvector[j] \Bigg) X^i , \quad \forall t \in \interval{\noseq+1}.
\end{align*}
Define an error-locator polynomial in $\Fxsub{q^{\QCCext}}$:
\begin{equation} \label{eq_ELP}
\Lambda(X) = \sum_{i=0}^{\noerrors} \Lambda_i X^i \defeq \prod_{i \in \errorsup}(1-X\alpha^{i \HTmult}).
\end{equation}
Like in the classical case of cyclic codes, we get $\noseq+1$ \textit{Key Equations} with a common error-locator polynomial $\Lambda(X)$ as defined in~\eqref{eq_ELP}:
\begin{equation} \label{eq_KeyEquation}
\Lambda(X) \cdot S_t(X) \equiv \Omega_t(X) \mod X^{\lenseq-1}, \quad \forall t \in \interval{\noseq+1},
\end{equation}
where the degree of each of $\Omega_0(X), \Omega_1(X), \dots, \Omega_{\noseq}(X) $ is smaller than $\noerrors$.
Solving these $\noseq+1$ Key Equations~\eqref{eq_KeyEquation} jointly can be realized by multi-sequence shift-register synthesis and several efficient realizations exist~\cite{feng_generalized_1989, feng_generalization_1991, zeh_fast_2011}.

Solving~\eqref{eq_KeyEquation} jointly is equivalent to solving the following heterogeneous system of equations: 
\begin{equation} \label{eq_DecodingSystem}
\begin{pmatrix}
\mathbf{S}^{\langle 0 \rangle} \\
\mathbf{S}^{\langle 1 \rangle} \\
\vdots \\
\mathbf{S}^{\langle \noseq \rangle}
\end{pmatrix}
\begin{pmatrix}
\Lambda_{\noerrors} \\
\Lambda_{\noerrors-1} \\
\vdots \\
\Lambda_{1}
\end{pmatrix}
=
\begin{pmatrix}
\mathbf{T}^{\langle 0 \rangle} \\
\mathbf{T}^{\langle 1 \rangle} \\
\vdots \\
\mathbf{T}^{\langle \noseq \rangle}
\end{pmatrix},
\end{equation}
where each $(\lenseq-1-\noerrors) \times \noerrors$ submatrix is a Hankel matrix:
\begin{equation} \label{eq_SyndromeMatrix}
\mathbf{S}^{\langle t \rangle} = \big ( S^{\langle t \rangle}_{i+j} \big)_{i \in \interval{\lenseq-1-\noerrors}}^{j \in \interval{\noerrors}}
, \quad \forall t \in \interval{\noseq+1},
\end{equation}
and each $\mathbf{T}^{\langle t \rangle} = (S_{\noerrors}^{\langle t \rangle} \ S_{\noerrors+1}^{\langle t \rangle} \ \dots \ S_{\lenseq-2}^{\langle t \rangle})^T$ with:
\begin{align*}
S_i^{\langle t  \rangle} & = \sum_{j=0}^{\QCCcyc-1} r_j(\alpha^{\HTconst + i \HTmult + t}) \eigenvector[j], \quad \forall i \in \interval{\lenseq-1}, t \in \interval{\noseq+1}.
\end{align*}

\begin{theorem}[Decoding up to New Bound] \label{theo_DecodingHTLikeBound}
Let $\QCC$ be an $\QCCcyc$-quasi-cyclic code and let the conditions of Thm.~\ref{theo_HTBound} hold. Let \eqref{eq_DecRadius} be fulfilled, let the $\noseq +1$ syndrome polynomials $S_0(X), S_1(X), \dots, S_{\noseq}(X)$ be defined as in~\eqref{eq_DefSyndromes}, and let the set of burst errors$\errorsup = \{j_0,j_1,\dots,j_{\noerrors-1} \}$ be as defined in~\eqref{eq_ModError}.
Then, the syndrome matrix $\mathbf{S} = (\mathbf{S}^{\langle 0 \rangle} \ \mathbf{S}^{\langle 1 \rangle} \ \dots \ \mathbf{S}^{\langle \noseq \rangle} )^T$ with the submatrices from~\eqref{eq_SyndromeMatrix} has $\rank(\mathbf{S})=\noerrors$.
\end{theorem}
\begin{IEEEproof}
Assume w.l.o.g. that $\HTconst = 0$. Similar to \cite[Section~VI]{feng_generalization_1991}, we can decompose the syndrome matrix into three matrices as follows: $\M{S} = (\M{S}^{\langle 0 \rangle} \ \M{S}^{\langle 1 \rangle} \ \cdots \ \M{S}^{\langle \noseq \rangle} )^T = \M{X} \cdot \M{Y} \cdot \overline{\M{X}} = ( \M{X}^{\langle 0 \rangle} \ \M{X}^{\langle 1 \rangle} \ \cdots \ \M{X}^{\langle \noseq \rangle})^T \cdot \M{Y} \cdot \overline{\M{X}}$,
where $\M{X}$ is a $(\noseq+1)(\lenseq-1-\noerrors) \times \noerrors$ matrix over $\F{q^{\QCCext}}$ and $\M{Y}$ and $\overline{\M{X}}$ are $\noerrors \times \noerrors$ matrices over $\F{q^{\QCCext}}$. Explicitly the decomposition provides the following matrices:
\begin{align*}
\M{X}^{\langle t \rangle} & = \big( \alpha^{(t+\HTmult i )j} \big)_{i \in \interval{\lenseq - 2 - \noerrors}}^{j \in \errorsup}, \quad t \in \interval{\noseq+1},\\
\overline{\M{X}} & = \big( \alpha^{i \HTmult j } \big)_{i \in \errorsup}^{j \in \interval{\noerrors}}, \quad \M{Y}  = \diag(E_{i_0},E_{i_1},\dots,E_{i_{\noerrors-1}}),
\end{align*}
where $E_{i} \defeq \sum_{t=0}^{\QCCcyc-1} e_{i,t} \eigenvector[t]$ for all $i \in \errorsup$.

Since $\M{Y}$ is a diagonal matrix, it is non-singular. From $\gcd(\QCClen, \HTmult) = 1$, we know that $\overline{\M{X}}$ is a Vandermonde matrix and has full rank. 
Hence, $\M{Y} \cdot \overline{\M{X}}$ is a non-singular $\noerrors \times \noerrors$ matrix and therefore $\rank(\M{S}) = \rank(\M{X})$. 
In order to analyze the rank of $\M{X}$, we proceed similarly as in \cite[Sec.~VI]{feng_generalization_1991}.
We use the matrix operation from \cite{van_lint_minimum_1986} % (see Corollary~\ref{coro_MatrixRank} in the appendix) 
to rewrite $\M{X} = \M{A} * \M{B}$, where
\begin{equation*}
\M{A} = \big( \alpha^{ij} \big)_{i \in \interval{\noseq+1}}^{j \in \errorsup} \quad \text{and} \quad \M{B} = \M{X}^{\langle 0 \rangle}.
\end{equation*}
We know from~\cite{van_lint_minimum_1986} that, if $\rank(\M{A}) + \rank(\M{B}) > \noerrors$, then $ \rank(\M{A} * \M{B}) = \noerrors$. Since $\gcd(\QCClen, \HTmult) =1$, both matrices $\M{A}$ and $\M{B}$ are Vandermonde matrices with $\rank(\M{A}) = \min\{\noseq+1, \noerrors \}$ and $ \rank(\M{B}) = \min\{\lenseq-1-\noerrors, \noerrors \}$. Assume w.l.o.g. that $(\lenseq -1) > \noseq$ (else we can interchange the roles $\lenseq$ and $\noseq$ in Thm.~\ref{theo_HTBound}). Therefore, from~\eqref{eq_DecRadius} we obtain $ \noerrors \leq (d^{\ast}-1)/2 = (\lenseq + \noseq -1)/2 < \lenseq-1$. Hence, investigating all four possible cases of $\rank(\M{A}) + \rank(\M{B})$ gives:
\begin{align*}
 & \noseq+ 1 + \lenseq - 1 - \noerrors  \geq 2 \noerrors - \noerrors +1 = \noerrors +1 > \noerrors,\\
 & \noseq + 1 + \noerrors > \noerrors, \\
 & \noerrors + \lenseq-1-\noerrors = \lenseq -1 > \noerrors,\\
 & \noerrors + \noerrors = 2\noerrors  > \noerrors,
\end{align*}
Thus, $\rank(\M{A}) + \rank(\M{B}) > \noerrors$.
\end{IEEEproof}
Algorithm~\ref{algo:decalgo} summarizes the whole decoding procedure, where the complexity is dominated by the operation in Line~\ref{algo_KeyEquation}. After the syndrome calculation (in Line~\ref{algo_SyndCalc} of Algorithm~\ref{algo:decalgo}), the $\noseq+1$ Key Equations~\eqref{eq_KeyEquation} are solved jointly (here in Line~\ref{algo_KeyEquation} with a Generalized Extended Euclidean Algorithm, GEEA~\cite{feng_generalized_1989}). Various other algorithms for solving the Key Equations jointly as in Line~\ref{algo_KeyEquation} with sub-quadratic time complexity exist.
 Afterwards, the roots of $\Lambda(X)$ as defined in~\eqref{eq_ELP} correspond to the positions of the burst errors as defined in~\eqref{eq_ModError} (see Line~\ref{algo_RootFinding}).

The error values $E_{i_0}, E_{i_0}, \dots E_{i_{\noerrors-1}}$ can be obtained from one of the $\noseq+1$ polynomials $\Omega_j(X)$ as given from the Key Equations~\eqref{eq_KeyEquation} (see Line~\ref{algo_BigErrorValues} in Algorithm~\ref{algo:decalgo}). In Line~\ref{algo_SmallErrorValues}, each error value $E_{i_j} \in \F{q^{\QCCext}}$ is mapped back to the $\QCCcyc$ error symbols $e_{i_j,0}, e_{i_j,1}, \dots, e_{i_j,\QCCcyc-1} \in \Fq$ and the codeword $\mathbf{c}(X) = (c_0(X) \ c_1(X) \ \dots \ c_{\QCCcyc-1}(X) )$ can be reconstructed.
\printalgo{\renewcommand{\algorithmcfname}{Algo}
\caption{\textsc{Decoding an $\LINQCC{\QCClen}{\QCCcyc}{\QCCk}{\QCCd}{q}$ Quasi-Cyclic Code}}
\label{algo:decalgo} 
\DontPrintSemicolon 
\SetAlgoVlined
\LinesNumbered
\SetKwInput{KwIn}{Input}
\SetKwInput{KwOut}{Output}
\BlankLine
\KwIn{Parameters $\QCClen, \QCCcyc,k, q, \QCCext$ of the quasi-cyclic code\\
Received word $\mathbf{r}(X) = (r_0(X) \ \dots \ r_{\QCCcyc-1}(X)) \in \Fqx^{\QCCcyc}$\\
Integers $\HTconst, \lenseq > 2, \noseq \geq 0$ and $\HTmult > 0$ with $\gcd(\HTmult,\QCClen)=1$\\
Eigenvalues $\eigenvalue{i} = \alpha^{\HTconst+i \HTmult+j}, \; \forall i \in \interval{\lenseq-1}, j \in \interval{\noseq+1} $\\
Eigenvector $(\eigenvector[0] \ \eigenvector[1] \ \dots \ \eigenvector[\QCCcyc-1]) \in \F{q^{\QCCext}}^{\QCCcyc}$}
\KwOut{Estimated codeword\\
$ \mathbf{c}(X)=(c_0(X) \ c_1(X) \ \dots \ c_{\QCCcyc-1}(X) ) $ \\
or \textsc{Decoding Failure}}
\BlankLine 
\BlankLine 
Calculate $S_0(X),S_1(X),\dots, S_{\noseq}(X)$ as in~\eqref{eq_DefSyndromes} \nllabel{algo_SyndCalc} \; 
\BlankLine
Solving Key Equations jointly $(\Lambda(X), \Omega_0(X), \Omega_1(X), \dots, \Omega_{\noseq}(X))$ = 
\texttt{GEEA$\big( X^{\lenseq-1}, S_0(X), S_{1}(X), \dots, S_{\noseq}(X) \big)$} \nllabel{algo_KeyEquation}\; 
\BlankLine
Find all $i$: $\Lambda(\alpha^{-i \HTmult})=0$ $\Rightarrow$ ${\mathcal E}=\lbrace i_0,i_1,\dots,i_{\noerrors-1}\rbrace$ \nllabel{algo_RootFinding}\; 
\BlankLine
\If{$\noerrors < \deg \Lambda(X)$}
{Declare \textsc{Decoding Failure}}
\Else{Determine error values $E_{i_0},E_{i_1},\dots,E_{i_{\noerrors-1}} \in \F{q^{\QCCext}}$ \nllabel{algo_BigErrorValues} \;
Determine $e_{i_j,0}, e_{i_j,1}, \dots, e_{i_j,\QCCcyc-1} \in \Fq$, s.t. $\sum_{t=0}^{\QCCcyc-1} e_{i_j,t}v_t = E_{i_j}, \quad \forall i_j \in \mathcal E$ \nllabel{algo_SmallErrorValues}\;
\BlankLine
$e_i(X) \leftarrow \sum_{j \in \mathcal{E}_i} e_{j,i} X^{j}, \quad \forall i \in \interval{\QCCcyc}$ \;
\BlankLine
$c_i(X) \leftarrow r_i(X) - e_i(X), \quad \forall i \in \interval{\QCCcyc}$\;}
}

\begin{example}[Decoding up to HT-like New Bound] \label{ex_DecodingBound}
Suppose the all-zero codeword of the $\LINQCC{63}{2}{100}{6}{2}$ $2$-quasi-cyclic code from Example~\ref{ex_HTbound} was transmitted. Let the two received polynomials in $\Fxsub{2}$ be:
\begin{equation*}
r_0(X) = e_0(X) = 1+X^{32}, \quad \quad r_1(X) = e_1(X) = X^{32}.
\end{equation*}
We have $\tilde{\noerrors} = 3$, but $\noerrors = 2$ (see~\eqref{eq_ModError}). The eigenvector $\eigenvector^{(5)} = (1 \ \alpha^4+1) \in \F{2^6}^2$ is contained in the intersection of the eigenspaces $\cap_{i \in D} \eigenspace[i]$, where $D \defeq \{0,4,8,1,5,9\} $, and is used for decoding. The system of two equations as in~\eqref{eq_DecodingSystem} becomes here:
\begin{equation*}
\begin{pmatrix}
\alpha^{35}  & \alpha^{26} \\
\alpha^{45} & \alpha^{33}
\end{pmatrix}
\begin{pmatrix}
\Lambda_2\\
\Lambda_1
\end{pmatrix}
=
\begin{pmatrix}
\alpha^{7}\\
\alpha^{51}
\end{pmatrix},
\end{equation*}
and the corresponding error-locator polynomial is $\sum_{i=0}^{2} \Lambda_iX^i =1+\alpha^{49}X+\alpha^{2}X^2 = (1-X)(1-X \alpha^{128}) $. The error-evaluation gives the two error values in $\F{2^6}$:
$E_0 = 1$ and $E_{32} = \alpha^4$. Therefore we can reconstruct the $\tilde{\noerrors} = 3$ error values $e_{0,0} = 1$, $e_{32,0}=1$ and $e_{32,1}=1$ in $\F{2}$.
\end{example}

\section{Conclusion and Outlook} \label{sec_conclusion}
We proved a new lower bound on the minimum distance of quasi-cyclic codes based on the spectral analysis introduced by Semenov and Trifonov. 
Moreover, a syndrome-based decoding algorithm was developed and its correctness proven.

\section*{Acknowledgments}
This work was initiated when S. Ling was visiting the CS Department of the Technion. He thanks this institution for its hospitality.

\vspace{-.3cm}

\printbibliography
\end{document}